\documentclass[conference]{IEEEtran}
\IEEEoverridecommandlockouts
\usepackage{cite}
\usepackage{amsmath,amssymb,amsfonts}
\usepackage{graphicx}
\usepackage{textcomp}
\usepackage{pgffor}
\usepackage{longtable}
\usepackage{multirow}
\usepackage{graphicx}
\usepackage{array}  
\usepackage{subcaption}
\usepackage{booktabs}
\usepackage{placeins} 
\usepackage{soul}
\usepackage{algpseudocode}
\usepackage{algorithm}
\usepackage{xcolor}
\setstcolor{red}

\usepackage{tabularx}
\usepackage{booktabs}
\usepackage{array}
\usepackage{enumitem}
\usepackage{verbatim}
\usepackage{listings}

\def\BibTeX{{\rm B\kern-.05em{\sc i\kern-.025em b}\kern-.08em
    T\kern-.1667em\lower.7ex\hbox{E}\kern-.125emX}}
\begin{document}

\title{Integrating Generative AI into Cybersecurity Education: A Study of OCR and Multimodal LLM-assisted Instruction
}

\author{
\IEEEauthorblockN{
Karan Patel\IEEEauthorrefmark{2},
Yu-Zheng Lin\IEEEauthorrefmark{2},
Gaurangi Raul\IEEEauthorrefmark{1},
Bono Po-Jen Shih\IEEEauthorrefmark{5},
Matthew W. Redondo\IEEEauthorrefmark{2},\\
Banafsheh Saber Latibari\IEEEauthorrefmark{2},
Jesus Pacheco\IEEEauthorrefmark{4},
Soheil Salehi\IEEEauthorrefmark{2},
Pratik Satam\IEEEauthorrefmark{2}\IEEEauthorrefmark{3}
}

\IEEEauthorblockA{
\IEEEauthorrefmark{1}College of Information Science, University of Arizona, Tucson, AZ, USA\\
\IEEEauthorrefmark{2}Department of Electrical and Computer Engineering, University of Arizona, Tucson, AZ, USA\\
\IEEEauthorrefmark{3}Department of Systems and Industrial Engineering, University of Arizona, Tucson, AZ, USA\\
\IEEEauthorrefmark{4}Department of Industrial Engineering, University of Sonora, Hermosillo, Mexico\\
\IEEEauthorrefmark{5}Leonhard Center for Enhancement of Engineering Education, The Pennsylvania State University, University Park, PA, USA\\
}
\IEEEauthorblockA{
Emails: \{graul\IEEEauthorrefmark{1}, yuzhenglin\IEEEauthorrefmark{2}, karanpatel\IEEEauthorrefmark{2},
mredondo245\IEEEauthorrefmark{2},
banafsheh\IEEEauthorrefmark{2},
ssalehi\IEEEauthorrefmark{2}, pratiksatam\IEEEauthorrefmark{2}\IEEEauthorrefmark{3}\}@arizona.edu;\\
jesus.pacheco@unison.mx\IEEEauthorrefmark{4}; bps5848@psu.edu\IEEEauthorrefmark{5}
}
}

\maketitle

\begin{abstract}
This research category full paper describes an LLM assisted instruction integration with a virtual cybersecurity lab
platform. The rapid digital transformation of Fourth Industrial Revolution (4IR) systems is transforming today's workforce needs, increasing skill set gaps, especially for the older workforce. With an increasing emphasis on STEM skill sets like robotics, automation, artificial intelligence (AI), and security, the workforce will have to be re-skilled and up-skilled to meet future industry needs. Generative Artificial Intelligence (AI), has the potential to aid building of this workforce, by assisting trainee's by acting as instructional assistant, helping students build skill sets while performing experiential learning exercises.
This paper presents a generative AI based instructional assistant integrated into a prior experiential learning platform. This instructional assistant uses a zero-shot, optical character recognition (OCR) based Large Language Model (LLM) assisted Instruction pipeline integrated into the legacy Cybersecurity Labs-as-a-Service (CLaaS) platform developed in 2015. The instructional agent assits the experiential learning by extracting the text from slide images using Tesseract OCR and to generate simplied instructions by prompting a general-purpose LLM enabling real-time instruction optimization with minimal infrastructure and integration effort. The instructional assistant was evaluated in a live university course and where in student feedback was collected using a 1-to-10 rating scale, with an average feedback score of 7.83 on 42 responses, indicating a strong perceived usefulness of the instructional assistant. In addition, we conducted a comparative study using multimodal LLMs that directly interpret slide images with the OCR pipeline, to analyze the computational overhead, and operational costs. While the multi-modal LLM pipeline performed better on slides with dense visual content, the OCR-LLM-based pipeline delivered comparable pedagogical value on text-centric slides with a significantly lower overhead, and operational costs. This work demonstrates that a lightweight and easily integrable pipeline can effectively bring modern generative AI capabilities into legacy platforms, offering a scalable enhancement for improving student comprehension in technical education settings.
\end{abstract}

\begin{IEEEkeywords}
Cybersecurity education, Large Language Models, AI in education, adaptive learning, OCR text summarize, Generative AI, Agentic AI.
\end{IEEEkeywords}

\section{Introduction}

The Fourth Industrial Revolution (4IR) is transforming industries through the convergence of advanced technologies such as artificial intelligence (AI), cloud computing, and cyber-physical systems \cite{lin2024ddd,lin2023dt4i4}. This transformation presents a dual challenge for the education sector: to prepare a new generation of students for emerging roles and to reskill or upskill the existing workforce to keep pace with rapidly evolving demands. As of 2025, these challenges have become more pressing. The World Economic Forum projected that by this year, 50\% of all employees would need reskilling, with 40\% requiring training lasting up to six months to adapt to new technologies and job functions \cite{wef2020futurejobs}. In parallel, studies indicate that two-thirds of today’s core skills have shifted in importance, with nearly one-third of the required competencies now involving entirely new areas not traditionally emphasized in standard curricula \cite{li2024reskilling}. These findings highlight the scale and urgency of creating accessible, adaptable educational systems that support both workforce development and continuous learning.

One of the most critical areas impacted by this shift is STEM education, which plays a foundational role in equipping learners with the technical and analytical skills required in a rapidly evolving job market \cite{nsb2020vision}. Disciplines such as cybersecurity, which are essential to the 4IR landscape, often present steep learning curves due to their technical depth and abstract concepts. Compounding this challenge is the fact that students vary widely in how they learn. Differences in background knowledge, pace of learning, and cognitive preferences can significantly affect educational outcomes, wherein conventional instructional methods are not always sufficient to meet this diversity in learning needs \cite{du2024personalized,lin2025personalized}.

To address these challenges, this paper builds upon the Cybersecurity Lab as a Service (CLaaS) platform, originally introduced in \cite{tunc2015claas,satam2023cps}, which provides a scalable, cloud-based environment for delivering hands-on cybersecurity training. CLaaS uses Amazon Web Services (AWS) to deploy virtual machines (VMs) that emulate real-world scenarios, enabling trainees to perform practical security experiments via a web interface. The platform supports a variety of cybersecurity tasks including DDoS attacks, buffer overflows, packet sniffing, and ransomware attacks. In addition to its technical infrastructure, CLaaS includes an instructional slide deck that delivers experiment guidelines and theoretical context directly within the training interface, allowing students to access learning resources alongside the labs.

This work explores the use of generative AI to personalize experiential learning in real time, addressing the diverse needs of students in complex STEM domains, with the aim of personalizing the experiential learning instructions for a student's needs. The core objective is to enhance student comprehension by dynamically adapting instructional content to match varying levels of prior knowledge and learning pace. To realize this vision, we extend the CLaaS platform by integrating a large language model (LLM) into its instructional interface. The integration operates in a zero-shot setting, requiring no domain-specific fine-tuning or preprocessing, and enables the system to simplify slide content on demand based on student interaction.

In addition to this integration, the paper presents a second experiment exploring an alternative method for enabling generative AI functionality within educational platforms. Instead of using multimodal LLMs capable of interpreting visual inputs directly, we apply an optical character recognition (OCR) tool to extract text from instructional slide images and submit the raw, unprocessed text to a language model in a zero-shot setting. By evaluating performance across a variety of slide types, we demonstrate that this approach offers comparable results to vision-language models, while significantly reducing implementation complexity and computational cost. This experiment highlights a practical and cost-effective pathway for incorporating generative AI into existing educational systems without requiring access to resource-intensive multimodal architectures.

Preliminary deployment has shown positive student engagement and feedback, suggesting that LLM-assist can meaningfully contribute to improved comprehension and accessibility in engineering education.

      
\begin{figure}[ht]
    \centering
    \includegraphics[width=\linewidth]{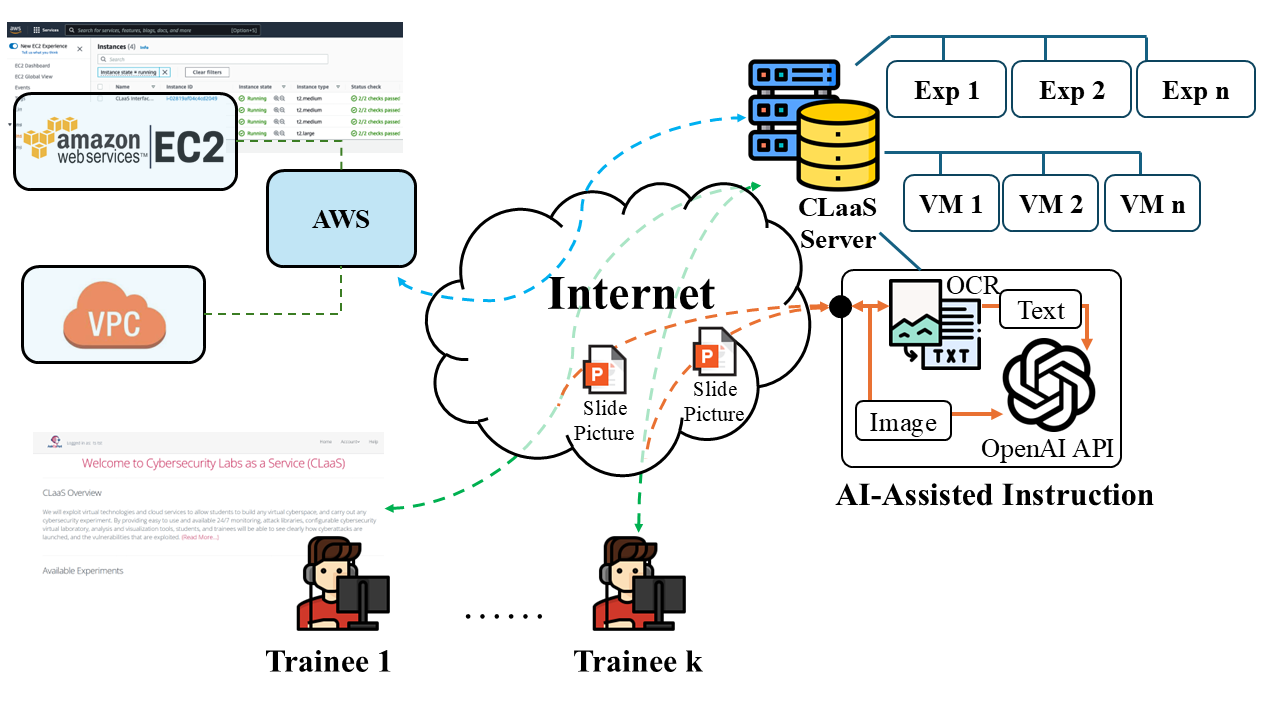}
    \caption{Generative AI based Instructional Assistant with Cybersecurity Labs as a Service (CLaaS) Architecture.}
    \label{fig:framework}
\end{figure}

\section{Related Work}

Generative AI encompasses a class of computational methods designed to produce novel and meaningful content such as text, images, or audio based on patterns learned from training data \cite{feuerriegel2024generative}. A prominent subset of this technology, large language models (LLMs), specializes in understanding and generating human language with high contextual fidelity. These models are trained on vast and diverse textual corpora, enabling them to support tasks involving reasoning, summarization, dialogue, and content generation \cite{fui2023generative}.
Recent advancements in LLMs have sparked increasing interest in their application within educational contexts \cite{kasneci2023chatgpt}, \cite{osunbunmi2024board}. Due to their ability to interpret language, generate explanations, and respond adaptively to prompts, LLMs are being integrated into intelligent tutoring systems, personalized learning platforms, and automated instructional design. For example, SimClass \cite{zhang2024simulating} leverages LLMs and multi-agent systems to simulate interactions in a virtual classroom. Evaluation across 400 students showed measurable improvements in engagement and learning outcomes, supporting the model's potential to enhance educational environments. Similarly, Chiang et al. \cite{chiang2023can} propose using LLMs as scalable evaluators in place of human raters for assessing natural language generation. Their findings indicate strong alignment between LLM-based evaluations and expert human judgments, particularly in dimensions such as fluency and relevance.
Other studies have explored discipline-specific applications. Tsai et al. \cite{tsai2023exploring} integrated LLMs into a chemical engineering course to support problem solving across core topics like thermodynamics and reaction engineering. Their experiment demonstrated notable gains in both conceptual understanding and problem-solving efficiency, particularly for tasks involving code and model generation.
Collectively, these studies illustrate the growing utility of LLMs in education. Their ability to offer scalable, interactive, and personalized assistance makes them well-suited to address challenges associated with traditional instruction.
While much of the prior work leverages fine-tuned models, our approach focuses on a lightweight, zero-shot integration of LLMs into an existing cybersecurity education platform. Specifically, we embed a generative language model into the platform’s instructional slide deck to simplify complex technical content in real time. This enables the delivery of adaptive explanations without requiring any domain-specific training or content preprocessing. By evaluating this integration in the context of cybersecurity education, we assess the effectiveness of zero-shot generative AI as a practical and scalable tool for improving student comprehension in STEM learning environments.\\
Lin et al. \cite{lin2024prism,lin2025personalized} have explored the usage of generative AI for experiential learning personalization, where in they have explored a Digital Twin based virtual reality (VR) learning environment \cite{alhamadah2024photogrammetry}. All student interactions in the VR environment are observed to judge the student skill set development mapped onto Blooms Taxonomy. A generative AI system, personalizes the experiential learning for each students learning needs based on observed student interactions, and the expected target skillset development on the Bloom's Taxonomy. These studies lay a strong foundation for the development of AI-assisted instruction. However, current applications still face limitations, especially in engineering education, where deeper integration with domain-specific tools and instructional systems remains challenging. Furthermore, the development of effective prompts and the integration of domain-specific input for LLM agents present ongoing difficulties, particularly for novice learners within a novel subject area. Even the competence to engage productively with LLMs in itself represents a significant learning hurdle, demanding meta-cognitive and self-regulated learning skills \cite{Urban2025BJET, Xu2025BJET}. These challenges underscore the necessity for instructors to implement instructional scaffolding by deliberately establishing the learning context within an LLM-assisted instructional system.

\section{Generative AI based Cybersecurity Labs as a Service (CLaaS) Architecture}


\begin{figure}[!t]
  \centering
  \fbox{\includegraphics[width=1\linewidth]{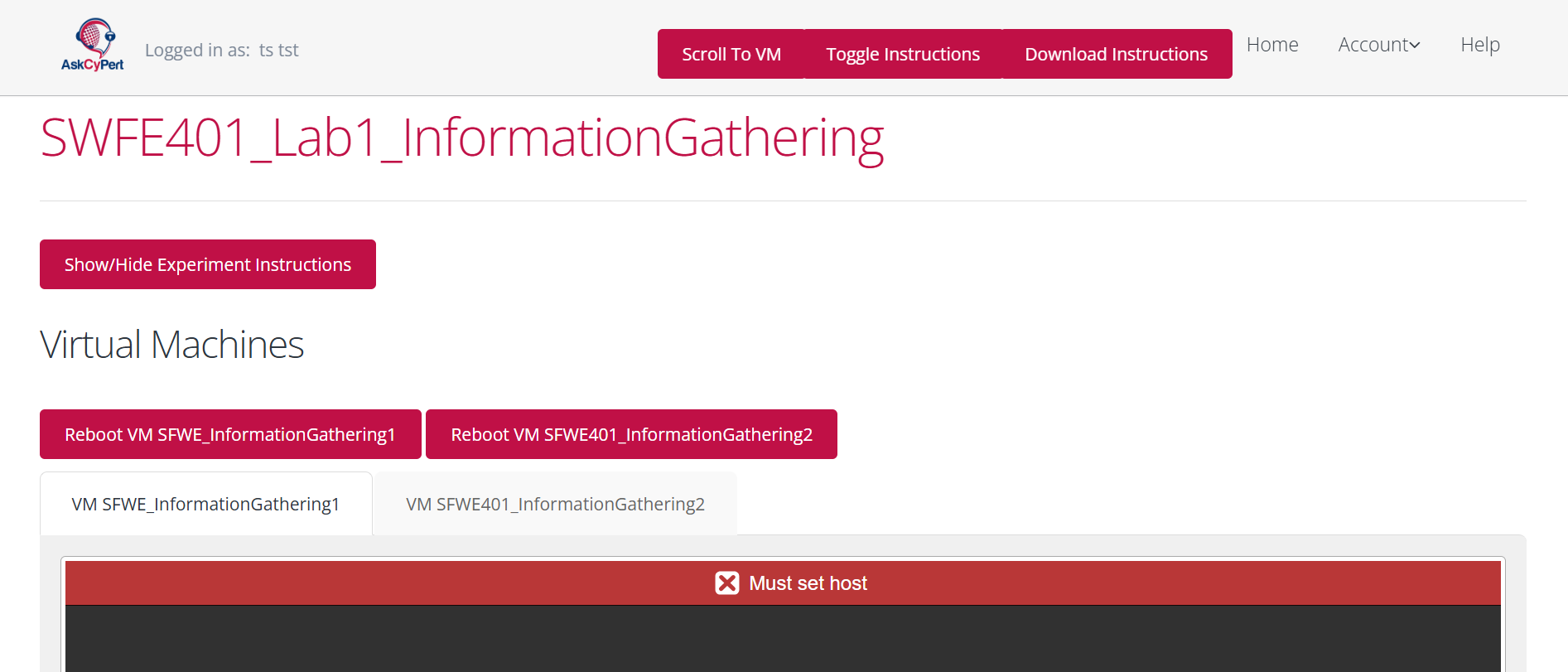}}
  \caption{ClaaS experimentation user interface.}
  \label{fig:vnc}
\end{figure}
This section describes the Cybersecurity Labs-as-a-Service (CLaaS) platform, with its generative AI agent based instruction personalization. CLaaS is a cloud-based software platform designed to deliver scalable, browser-accessible cybersecurity lab environments. \texttt{CLaaS} provisions isolated virtual machines for each student using pre-configured images tailored to specific experiments. Students connect to these environments via an embedded VNC interface as shown in Figure~\ref{fig:vnc}, enabling them to complete labs entirely within their web browser without the need for local setup. Within the platform, each lab includes a series of instructional slides and procedural steps that guide students through technical tasks. 

The proposed research extends the Cybersecurity Lab as a Service (CLaaS) platform by integrating a generative AI–driven instructional assistant designed to simplify technical content in real time. The enhanced architecture preserves the core functionality of CLaaS while introducing a new capability that supports adaptive content generation using large language models (LLMs). 

\subsection{CLaaS Architecture}
The CLaaS (as shown in Figure \ref{fig:framework}) is an AWS cloud-based web service that allows instructors to create cybersecurity training scenarios. The CLaaS web service, built on a client-server model, aims to reduce the instructor's load allowing training material reusability while providing the capability to include hands-on training into an online asynchronous curriculum. The CLaaS is hosted on an AWS EC2 instance server manages the complete training performing the following tasks: 
\subsubsection{User authentication and access control} 
CLaaS has three user roles: a) Administrator: The administrator manages the CLaaS and has complete access to the system and its functioning. The administrator can create other users in all three roles and access the CLaaS server backend through remote access; b) Instructor: A user in the instructor role has permission to manage a course and access CLaaS functionality to create new scenarios and assign them to trainees; c) Student: A user in the student role has least permissions in the CLaaS and can only execute scenarios assigned to them by the instructor.
\subsubsection{CLaaS experiment management}
The CLaaS uses AWS EC2 instances with VPC subnets for executing the training scenarios. The CLaaS server manages these scenarios by performing the following: 
\begin{enumerate}
    \item Scenario setup: During the scenario setup, the CLaaS creates new EC2 instances (using stored scenario images) and new subnets in a VPC and connects them to enable the scenario. The CLaaS server also setups CloudWatch alarms to manage and limit the EC2 instance resource utilization to limit the training range's operational costs; 
    \item Scenario startup and stopping: The CLaaS server, starts the EC2 instances for training scenarios when a user begins an experiment, and stops the instances when the user turns off the training scenario, reducing the operational costs of the server; 
    \item Scenario Termination: An instructor can terminate a scenario for a student, freeing all the resources (EC2 instances and subnet) assigned to that student. The CLaaS server allows the instructors and administrators to perform scenario termination.
\end{enumerate}

\subsubsection{CLaaS front-end hosting} The CLaaS has a web front-end for instructors and students to create and practice the scenarios, respectively. The front-end provides an interface to manage student enrollment, experiments, assignments, and experiment performance pages. The experiment performance page has a noVNC plugin that allows the front-end to embed a VNC connection to the EC2 instances enabling the trainee to perform different scenarios.

As part of this research, a generative AI based instructional assistant module is integrated into the CLaaS ecosystem as shown in figure \ref{fig:framework}. This module enables students to request simplified explanations of instructional slides directly within the platform. Slide content is captured as images and transmitted to the LLM-assist pipeline, which consists of an optical character recognition (OCR) layer followed by an interaction with an LLM via an API (e.g., OpenAI's GPT model). The extracted text is sent to the LLM without additional preprocessing, allowing the model to generate a simplified explanation of the slide content in a zero-shot manner. The response is then delivered to the student via the CLaaS interface.


\subsection{CLaaS Training Scenario Architecture}

Each CLaaS training Scenario consists of EC2 instances  (VMs) and an AWS VPC-based subnet for the local network communication. The AWS EC2 instances are t2-micro instances, each having a Ubuntu20.04 operating system, 1GB ram, and 8GB storage space. These EC2 instances are built from a primary EC2 instance image with an XFCE user interface and a VNC server setup for auto-start on boot. While creating a training scenario, the instructor installs tools and configures them on instances created from the primary EC2 instance image. Once the tools setup is complete, the instructor saves the EC2 instances as new images that are used to duplicate the scenario for the trainees on assignment. The VPC-based subnet allows the EC2 instances in a scenario to communicate with one another. The subnet connects to a VPC gateway allowing the EC2 instances to allow connection to the network (depending on the configuration of the subnet security rules).

\subsection{Generative AI based Instruction Assistant}
We integrated an ``Instruction Simplication'' button directly into the slide interface of the CLaaS platform, that triggers the generative AI agent to assist the student. When a student clicks this button, the text content is extracted from the current slide using an optical character recognition (OCR) tool --- \texttt{Tesseract}~\cite{smith2007overview}. The extracted text is then combined with a carefully crafted prompt and sent to a Large Language Model (LLM), which returns an LLM-assisted Instruction version of the content tailored to enhance student comprehension.

Students can rate the usefulness of the AI assisted Instruction content using a 1–10 feedback scale. This feedback is stored and can be analyzed to assess the impact and usefulness of the LLM-generated content and can be further used to personalize the prompt for the student.





\subsection{Benchmarking OCR-to-LLM Against Multimodal LLMs}
To assess the potential of more advanced generative models, we also conducted a parallel capability using multimodal LLMs capable of interpreting slide images directly. Multimodal Large Language Models (LLMs) are AI systems capable of processing and generating content across multiple data modalities, such as text, images, audio, and video, enabling them to perform tasks that involve both language and visual understanding. These models extend the capabilities of traditional LLMs by incorporating vision-language alignment techniques, allowing them to interpret visual inputs (e.g., diagrams, screenshots, photos) alongside natural language prompts. For example, \texttt{GPT-4V} (the vision-enabled variant of \texttt{GPT-4}) accepts images as input and can generate detailed explanations, descriptions, or reasoning based on both visual and textual cues \cite{yang2023dawn}. 

This allowed us to compare the performance of a lightweight, OCR-based method, which is low-cost and easy to deploy, with a resource-intensive, image-based model. The \texttt{CLaaS} platform currently implements the OCR pipeline in production, and student feedback has been actively collected to evaluate its effectiveness. This modular setup allows for future integration of additional LLM interfaces and personalization mechanisms as needed.

\setlength{\fboxsep}{0pt}
\setlength{\fboxrule}{0.5pt}

\begin{figure*}[t]
    \centering
    \begin{subfigure}[b]{0.3\textwidth}
        \fbox{\includegraphics[width=\linewidth]{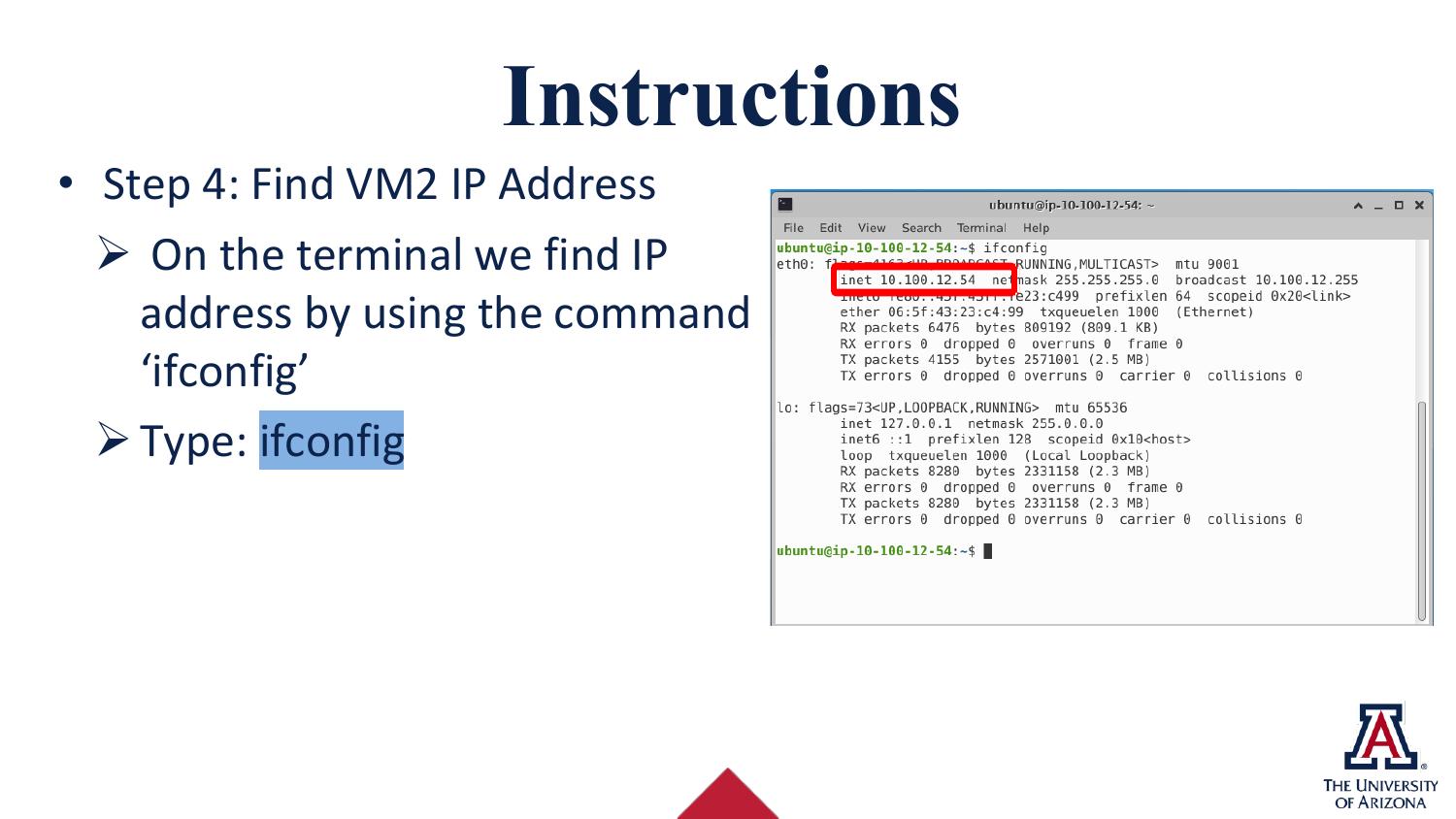}}
        \caption{Slide illustrating IP address retrieval using command-line interface with example output.}
        \label{fig:slide3}
    \end{subfigure}\hfill
    \begin{subfigure}[b]{0.3\textwidth}
        \fbox{\includegraphics[width=\linewidth]{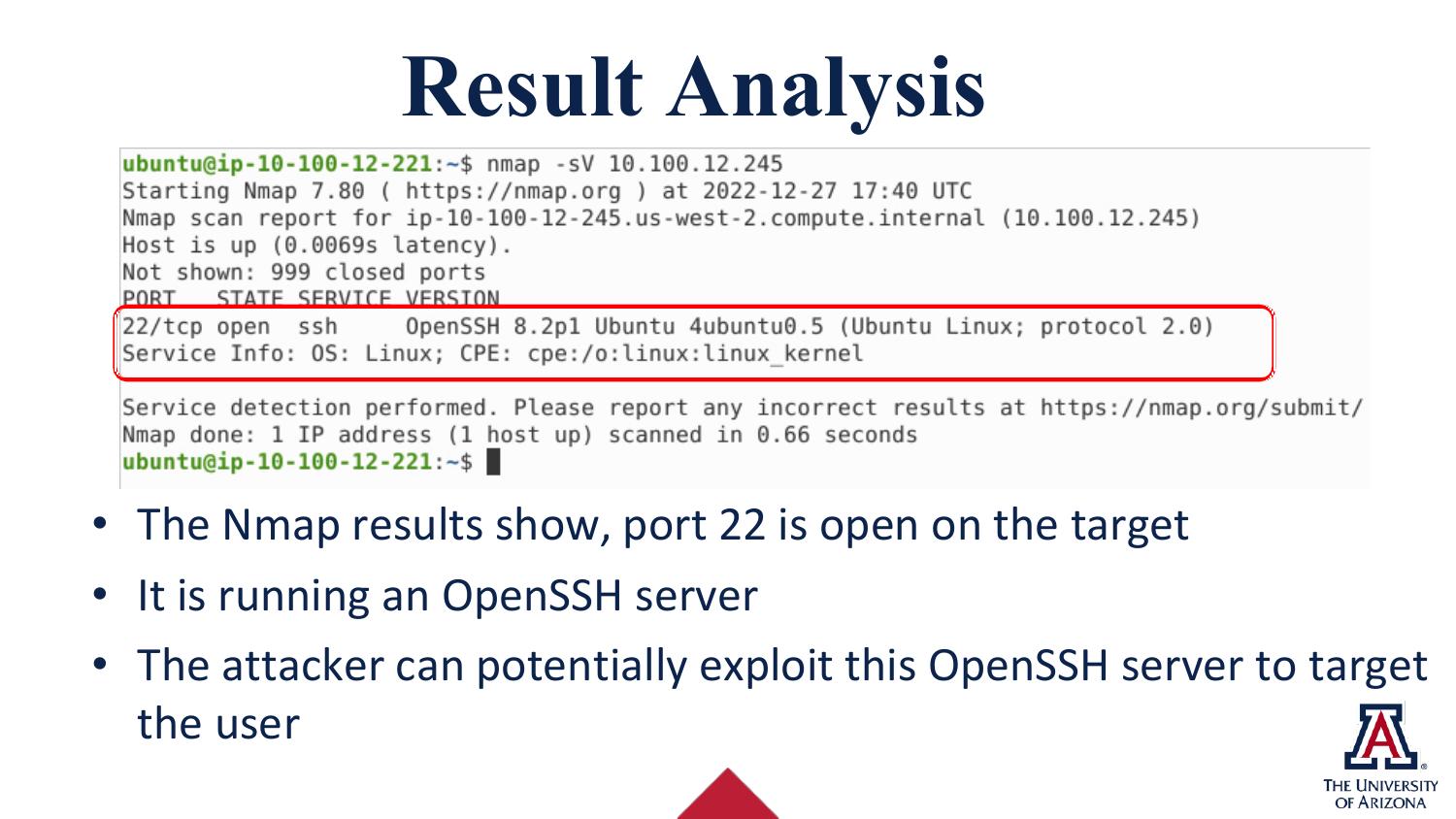}}
        \caption{Technical result slide summarizing Nmap port scanning analysis with structured bullet points.}
        \label{fig:slide1}
    \end{subfigure}\hfill
    \begin{subfigure}[b]{0.3\textwidth}
        \fbox{\includegraphics[width=\linewidth]{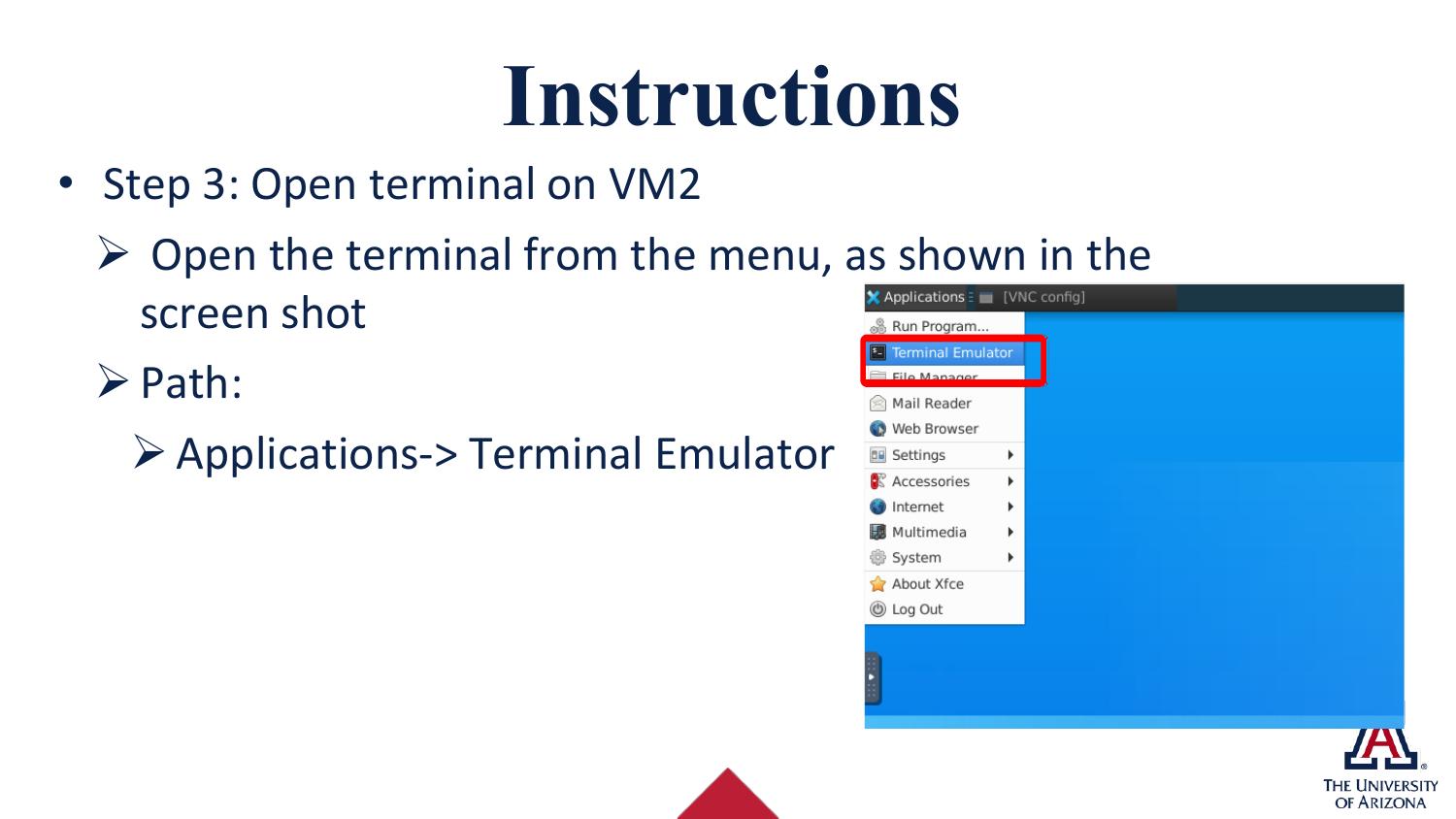}}
        \caption{Instructional slide demonstrating GUI-based terminal launch with visual highlights.}
        \label{fig:slide2}
    \end{subfigure}\\[1ex]

    \begin{subfigure}[b]{0.3\textwidth}
        \fbox{\includegraphics[width=\linewidth]{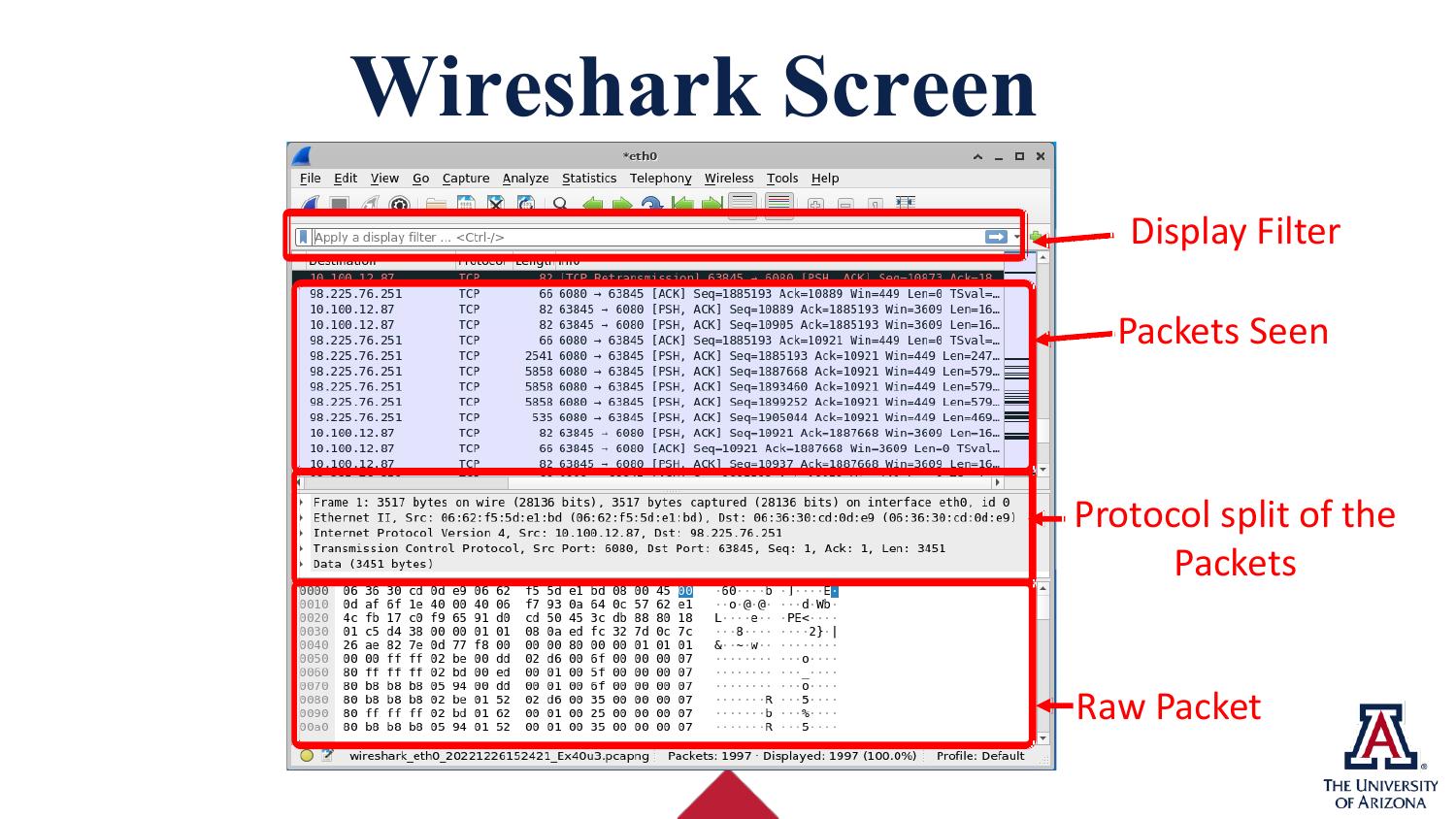}}
        \caption{Highly annotated slide showing Wireshark interface with labeled components for network packet inspection.}
        \label{fig:slide5}
    \end{subfigure}\hfill
    \begin{subfigure}[b]{0.3\textwidth}
        \fbox{\includegraphics[width=\linewidth]{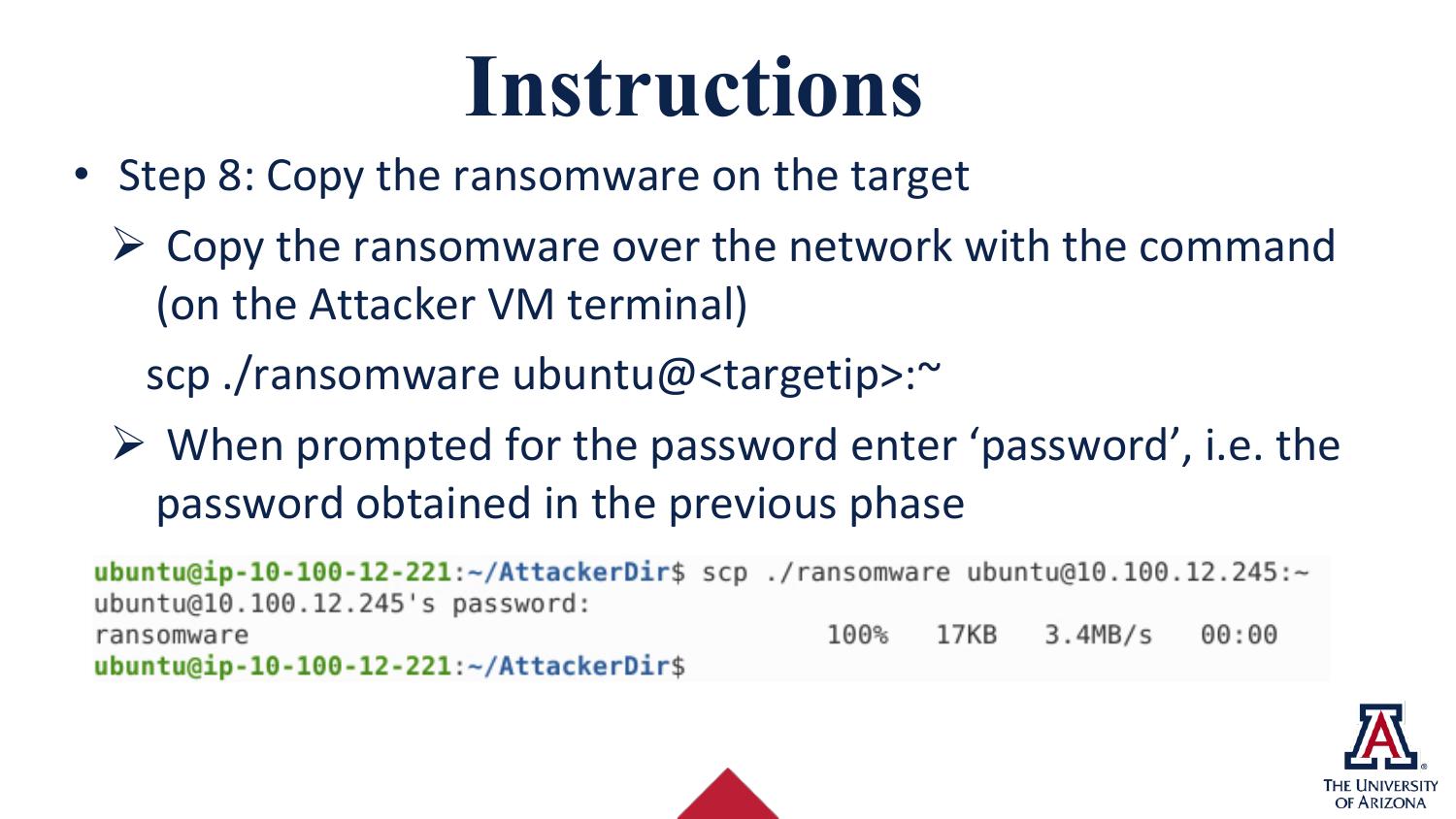}}
        \caption{Command-line focused slide demonstrating file transfer using \texttt{scp} with terminal outputs.}
        \label{fig:slide5}
    \end{subfigure}\hfill
    \begin{subfigure}[b]{0.3\textwidth}
        \fbox{\includegraphics[width=\linewidth]{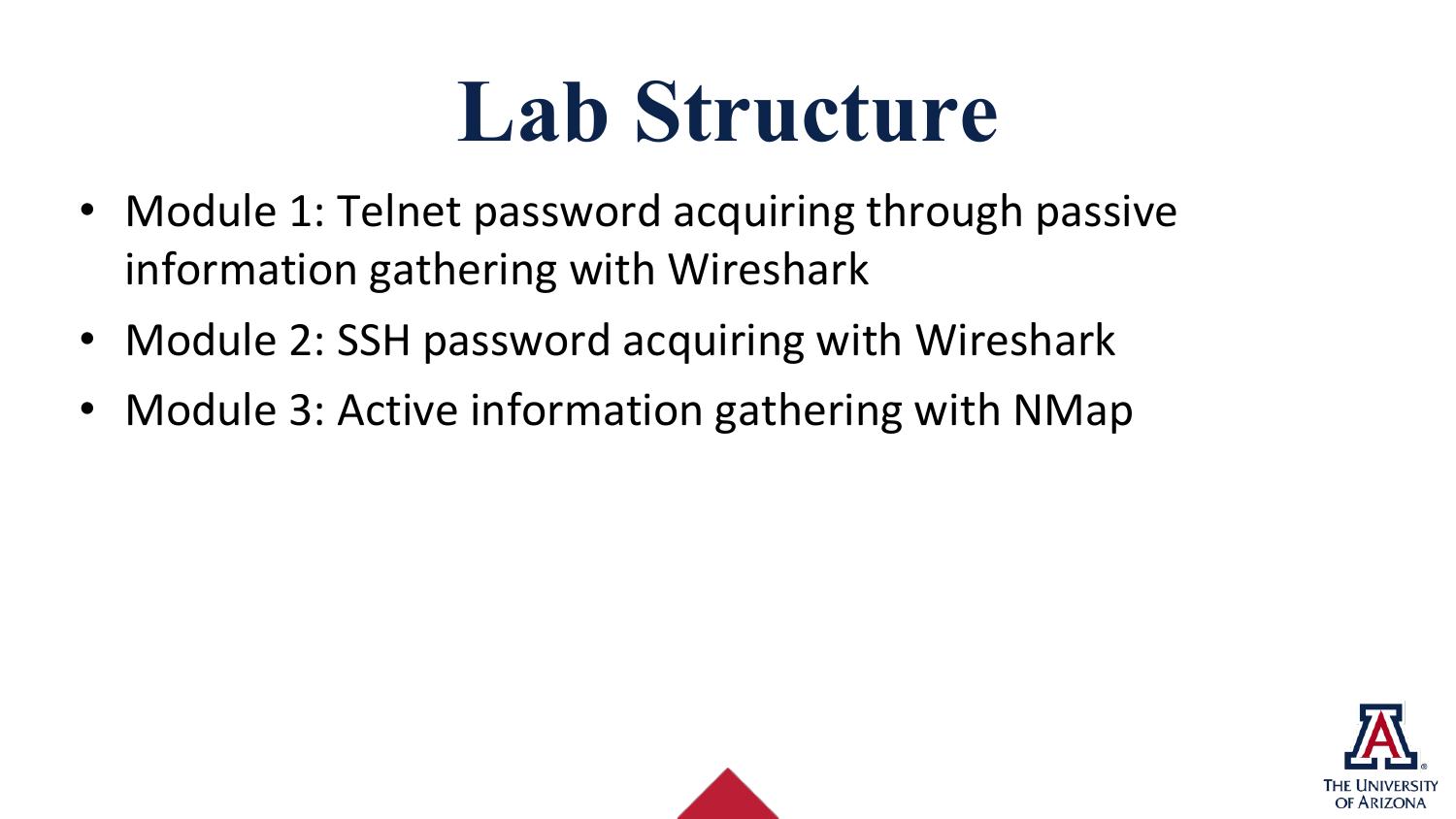}}
        \caption{Overview slide summarizing the lab modules in bullet format. (Note: Template also a kind of noise on OCR.)}
        \label{fig:slide11}
    \end{subfigure}

    \caption{Varying complexity of slide examples for LLM-assisted Instruction tasks.}
    \label{fig:six-slides}
\end{figure*}

\section{Experimental Results}

\subsection{Experimental setup}\label{AA} 
To evaluate the effectiveness of the generative AI based instruction personalization, the CLaaS environment was deployed in AWS where in, we conducted a two-part experiment comparing two methods:

\begin{itemize}
    \item OCR-to-LLM Pipeline (Text-Only): Slide content was extracted using \texttt{Tesseract} OCR and sent to an LLM (\texttt{gpt-4}) to generate LLM-assisted Instruction versions. This approach was integrated into the \texttt{CLaaS} platform and tested in a live university course, where students rated the usefulness of LLM-assisted Instruction using a 1–10 feedback scale.

    \item Multimodal LLM Pipeline (Image-Based): Slide images were fed directly into a multimodal LLM capable of interpreting visual content (\texttt{gpt-4o-mini}). This method was evaluated with Python script to compare the quality of instruction without requiring in-browser deployment.
\end{itemize}

The purpose of this experiment was to assess whether the lightweight, cost-effective OCR-based pipeline could produce instruction of clarity and usefulness comparable to those generated by more resource-intensive multimodal models.

The following cybersecurity experiential learning experiments were made available to students, and students were encouraged to use the AI assist feature and rate the effectiveness of simplified instruction on a scale of 1-10:\begin{itemize}
    \item \textbf{Experiment 1 --- Information Gathering}:  Students are introduced to passive and active information-gathering techniques commonly used in cybersecurity reconnaissance. Passive information gathering is performed using Wireshark, allowing students to observe and analyze network traffic without generating additional packets on the network. For active information gathering, students utilize Nmap to scan target systems, identify open ports, and run services through direct interaction with the network.

The slide materials for this lab include definitions, objectives, and expected outcomes. The AI-assist versions of these instructions and explanations to help students understand the tools and processes.

    \item \textbf{Experiment 2 --- Understanding Ransomware}:  Students will explore the complete lifecycle of a ransomware attack, gaining a deeper understanding of how such threats are orchestrated in real-world scenarios. The experiment is structured into three distinct phases:
\begin{itemize}
    \item \textbf{Phase 1}: Information Gathering – Students identify potential system weaknesses through reconnaissance techniques.

    \item \textbf{Phase 2}: Exploitation – Using the information collected, students simulate the exploitation of identified vulnerabilities to gain unauthorized access.

    \item \textbf{Phase 3}: Attack Delivery – Finally, students simulate the deployment of a ransomware payload to understand how such attacks are delivered and executed in a target environment.
\end{itemize}

\end{itemize}

\subsection{Generative AI Agent's performance evaluation via student feedback}

To evaluate the perceived usefulness of the OCR-based LLM-assisted Instruction feature, we collected direct feedback from undergraduate and graduate students enrolled in a university cybersecurity course. We collected 42 feedback ratings across various instructional slides. Students were prompted to rate each LLM-generated instruction on a scale of 1 to 10, reflecting how helpful they found the explanation in understanding the underlying slide content.

The average rating across all submissions was 7.83, indicating a generally positive reception of the generative AI agent's simplified instructions. This suggests that, even with occasional noise introduced by OCR artifacts, the generative AI agent was able to generate content that students found helpful and accessible.

\subsection{Comparative Evaluation of OCR-Based vs. Multimodal LLM Instruction}

To evaluate the effectiveness of the OCR-based text instruction pipeline against a multimodal LLM approach, we conducted a qualitative comparison across a series of instructional slides used in \texttt{CLaaS}. The slides varied in content, with some containing only instructional text and others including screenshots of terminal outputs or UI interfaces.

Key observations from the comparison include:

\begin{itemize}

\item Slides with Clean Text and Terminal Outputs (e.g., Figure 
 \ref{fig:six-slides} (a), Figure \ref{fig:six-slides} (b)):
The OCR-based approach was able to accurately extract both the slide text and terminal output. While the extracted-text LLM generated coherent instruction content, it sometimes included excessive or redundant verbiage due to the verbose OCR output. In contrast, image-based LLM produced more concise and focused explanations by interpreting the image context directly.

\item Slides with simple instructions and screen shots (see, e.g., Figure \ref{fig:six-slides} (c)):
OCR struggled to interpret visual elements and produced jumbled or noisy text, particularly from UI screenshots. However, despite the noisy input, the text-based LLM was still able to generate contextually accurate and helpful explanations, comparable to the image-based LLM's output.

\item Slides Combining Text Instructions with Annotated Screenshots (e.g., Figure \ref{fig:six-slides} (a) , Figure \ref{fig:six-slides} (e)):
In cases where the slide contained both textual instructions and illustrative screenshots (e.g., terminal commands or visual cues like red-circled areas), the OCR sometimes merged or misinterpreted content. Still, the text-based LLM often compensated for noisy input using domain knowledge, correctly identifying elements like IP address fields or key terminal outputs.

\item Slides with Minimal or No Text (e.g., Figure \ref{fig:six-slides} (d)):
For slides composed solely of annotated screenshots (e.g., Wireshark UI), the image-based LLM clearly outperformed. The OCR output was dense and inconsistent, leading to less precise instruction by the text-based LLM. Nonetheless, it was still able to identify keywords and provide partial understanding of the slide content.

\item Slides with Primarily Text-Based Content (e.g., Figure \ref{fig:six-slides} (f)):
For slides composed almost entirely of instructional text with minimal or no visual elements, both the OCR-based and image-based LLM approaches produced nearly identical instruction. This reinforces the viability of the OCR pipeline for slides where the visual context is not essential to understanding.

\end{itemize}

\begin{table}[]
\caption{LLM-assisted Instruction example}
\resizebox{\columnwidth}{!}{%
\begin{tabular}{c}
\hline
\textbf{LLM-assisted Instruction output for Figure \ref{fig:six-slides}(a)}                           \\ \hline
OCR Extracted Text                         \\ \hline
\multicolumn{1}{l}{\begin{tabular}[c]{@{}l@{}}Instructions\\ ° Step 4: Find VM2 IP Address\\ \textgreater On the terminal we find IP\\ address by using the command\\ ‘ifconfig’\\ \textgreater Type:\\ File Edit View Search Terminal Help\\ ubuntu@ip-10-100-12-54:\textbackslash{}textasciitilde\{\}\$ ifconfig\\ i UNNING ,MULTICAST\textgreater mtu 9061\\ :5:43:23:c4:99 txqueuelen 1000 (Ethernet)\\ RX packets 6476 bytes 809192 (899.1 KB)\\ RX errors 8 dropped @ overruns 0 frame 0\\ TX packets 4155 bytes 2571001 (2.5 MB)\\ TX errors 8 dropped @ overruns @ carrier @ collisions 6\\ lo: flags=73\textless{}UP,LOOPBACK,RUNNING\textgreater mtu 65536\\ inet 127.0.0.1 netmask 255.0.0.0\\ inet6 ::1 prefixlen 128 scopeid 6x19\textless{}host\textgreater\\ loop txqueuelen 1906 (Local Loopback)\\ RX packets 8280 bytes 2331158 (2.3 MB)\\ RX errors 8 dropped @ overruns 9 frame 0\\ TX packets 8280 bytes 2331158 (2.3 MB)\\ TX errors @ dropped @ overruns 9 carrier @ collisions 6\\ ubuntu@ip-10-106-12-54:\textbackslash{}textasciitilde\{\}\$ fj\\ ask 255.255.255.0 broadcast 10.100.12.255\\ 23:¢499 prefixlen 64 scopeid @x20\textless{}Link\textgreater\\ THE UNIVERSITY\\ OF ARIZONA\end{tabular}} \\ \hline
LLM-assisted Instruction (Text-based input)  \\ \hline
\multicolumn{1}{l}{\begin{tabular}[c]{@{}l@{}}Here's a simpler explanation of the text:\\ **Instructions**\\ **Step 4: Find the IP Address of VM2**\\ 1. You need to open the terminal to find the IP address.\\ 2. Use the command `ifconfig` to see the network configuration.\\ After typing the command, you might see something like this:\\ - The output includes various pieces of information about network connections.\\ - Look for details that show "inet" (which stands for Internet) followed by a number, \\ like `10.100.12.54`. This number is the IP address.\\ - There are also other numbers and information about data packets sent and received, \\ which you can ignore for this step.\\ So basically, this step tells you to use `ifconfig` in the terminal to find the IP address \\ of your VM2, focusing on the part that starts with "inet."\end{tabular}} \\ \hline
LLM-assisted Instruction (Image-based input) \\ \hline
\multicolumn{1}{l}{\begin{tabular}[c]{@{}l@{}}The image shows steps to find the IP address of a virtual machine (VM2). \\ 1. **Find the IP Address**: This is Step 4 of a series of instructions.\\ 2. **Use the Terminal**: You need to open a terminal window (a command line \\ interface) on your computer.\\ 3. **Type a Command**: You enter the command `ifconfig` to see your network \\ settings.\\ 4. **Look for the IP Address**: After typing the command, you will see various \\ information. Your IP address is shown next to "inet" (for example, `10.100.12.54`).\\ In simple terms, this is about checking your virtual machine's internet address \\ using a specific command.\end{tabular}} \\ \hline
\end{tabular}%
}
\label{tab:llm_output}
\end{table}

The Table \ref{tab:llm_output} is an example of OCR-based and Multimodal LLM comparison for Figure \ref{fig:six-slides} (a). Although the OCR result contains a lot of noise from terminal image recognition. However, LLM showed good anti-interference ability and extracted information from it as LLM-assisted Instruction. Although the OCR result contains a lot of noise from terminal image recognition. However, LLM showed good anti-interference ability and extracted information from it as LLM-assisted Instruction. Multimodal LLM directly reads images well and the response is more concise. In terms of cost, the OCR-based text input method in this example uses only 392 tokens for 1124 characters, but if image-based input is used to multi-modal LLM, an image file is $1500\times 844$ px, and the number of tokens used is 1105 tokens \cite{openai2025tokenizer,paxhumana2025tokenizer}. Therefore, the OCR-based text input method is more cost-effective. For the experiment results of Figure \ref{fig:six-slides} (b)-(f) is show in Appendix \ref{app:llm_result}.

Overall, the results indicate that while multimodal LLMs provide advantages for slides rich in visual content, the OCR-based pipeline is not far behind. It consistently produces accurate and helpful instruction across a wide range of slide types, particularly those that are text-dominant or structurally consistent. Given its low cost and ease of integration, the OCR pipeline offers a practical and scalable solution to bring modern LLM-based support to existing educational platforms. Since many instructional systems already rely heavily on structured content slides, the OCR approach enables a lightweight yet powerful enhancement to learner comprehension with minimal modification to existing infrastructure.  

\section{Conclusion and Future Work}
This paper presents a generative AI based instruction assistant that simplifies complex instructions for more effective experiential learning. The proposed generative AI agent was evaluated via deployment in a university course, with promising results from student feedback. Furthermore, our comparison with a multimodal LLM revealed that, while image-based models perform well on visually rich content, the OCR-based pipeline offers comparable instruction quality for text-centric slides with significantly lower integration overhead, making it a practical solution for real-world educational platforms.

The further study aims to extend this system with more advanced feedback analytics, including sentiment analysis. Plans are in place to generalize the system across multiple courses beyond cybersecurity. As the system continues to operate in \texttt{CLaaS}, we will collect and analyze more student data to refine instruction strategies and improve learning outcomes. In future iterations, we also plan to explore the personalization of instruction: adapting the output's tone, depth, or format based on individual learner preferences and past feedback.

Additionally, to enhance the accuracy of AI-assist output, the plan is to investigate using interactive or multi-turn LLM feedback, where students can ask follow-up questions on instruction content to simulate a conversational tutoring experience. Moreover, the system also considers exploring the integration of Retrieval-Augmented Generation (RAG) pipelines to further enhance personalization \cite{shan2024openrag}. This approach would allow the system to tailor explanations to the student’s feedback history and the exact terminology and examples used in the course, further improving coherence, alignment with curriculum objectives, and instructional clarity.

\section*{Acknowledgment}
This work was partially supported by the National Science Foundation (NSF) under research projects 2335046, the University of Arizona's Provost Investment Fund for AI Hardware Design League (AI-HDL), and the OpenAI Researcher Access Program 0000011862.

\bibliographystyle{IEEEtran} 
\bibliography{refs}

\appendices
\section{LLM-assisted Instruction results}\label{app:llm_result}

The LLM-assisted Instructions corresponding to Figures~\ref{fig:six-slides}(b)--(f) are presented in Tables~\ref{tab:slide2}--\ref{tab:slide6}, respectively.

\begin{table}[hbt!]
\centering
\caption{LLM-assisted Instruction output for Figure \ref{fig:six-slides}(b)}
\small
\begin{tabularx}{\columnwidth}{>{\raggedright\arraybackslash}X}
\toprule
\textbf{Text-based input with OCR} \\
\midrule
Here’s the text explained in simpler terms:
\begin{itemize}[leftmargin=*]
  \item A command was run on a computer to scan another computer's IP address (10.100.12.245) using a tool called Nmap.
  \item The scan started on December 27, 2022, and finished quickly in about 0.66 seconds.
  \item The scan found that the target computer is active (it's "up") and responding.
  \item It checked 1000 possible communication ports but most of them (999) are closed, which means they aren’t currently allowing connections.
  \item However, port 22 is open. This port is used for a service called OpenSSH, which allows secure remote access to the computer.
  \item The system running this service is Ubuntu Linux, which is a type of operating system.
  \item The Nmap tool also checks the type of service running and suggests reporting any mistakes in the results.
  \item Overall, the open port could be a security risk because attackers might exploit the OpenSSH service to target a specific user on that computer.
\end{itemize}
In summary, the scan showed that the target computer is reachable, has most of its ports closed, but is running a potentially vulnerable service at port 22. \\
\midrule
\textbf{Image-based input with multi-modal LLM} \\
\midrule
The image shows the results of a network scan using a tool called Nmap. Here’s a simpler explanation:
\begin{itemize}[leftmargin=*]
  \item \textbf{Nmap Scan}: The scan looks at a specific IP address (10.100.12.245) to see what services are running on it.
  \item \textbf{Port 22 is Open}: This means that port 22, which is commonly used for secure connections (SSH), is available for communication.
  \item \textbf{OpenSSH Server}: The server is using OpenSSH software to handle these secure connections.
  \item \textbf{Security Risk}: An attacker could take advantage of the open SSH service to try to gain unauthorized access to the user.
\end{itemize}
In summary, the scan shows that the target server has a security feature that could be vulnerable to attacks. \\
\bottomrule
\end{tabularx}
\label{tab:slide2}
\end{table}

\begin{table}[hbt!]
\centering
\caption{LLM-assisted Instruction output for Figure \ref{fig:six-slides}(c)}
\small
\begin{tabularx}{\columnwidth}{>{\raggedright\arraybackslash}X}
\toprule
\textbf{Text-based input with OCR} \\
\midrule
Here's a simpler version of the instructions:
\begin{itemize}[leftmargin=*]
  \item \textbf{Step 3: Open the Terminal on VM2}
  \item Find the terminal program on your computer by looking in the menu. You can refer to the picture for help.
  \item The menu path is: First go to “Applications”, then select “Terminal Emulator”.
\end{itemize}
That’s it! \\
\midrule
\textbf{Image-based input with multi-modal LLM} \\
\midrule
This image provides instructions on how to open a terminal on a virtual machine (VM2). Here's a simpler explanation:
\begin{itemize}[leftmargin=*]
  \item \textbf{Step 3}: You need to open a terminal.
  \item \textbf{How to Open}: Look at the menu on your screen as shown in the picture.
  \item \textbf{Where to Find It}: Go to the “Applications” menu and click on “Terminal Emulator” (it's highlighted in red).
\end{itemize}
That’s it! You just follow that path to find and open the terminal. \\
\bottomrule
\end{tabularx}
\label{tab:slide3}
\end{table}







\begin{table}[hbt!]
\centering
\caption{LLM-assisted Instruction output for Figure \ref{fig:six-slides}(d)}
\small
\begin{tabularx}{\columnwidth}{>{\raggedright\arraybackslash}X}
\toprule
\textbf{Text-based input with OCR} \\
\midrule
The text describes the display and information provided by Wireshark, a tool used to monitor and analyze network traffic. Here’s a simpler explanation:
\begin{itemize}[leftmargin=*]
  \item \textbf{Wireshark Menu}: The top part lists different menu options like File, Edit, View, Capture, Analyze, and Help. These allow users to manage files and analyze network data.
  \item \textbf{Basic Math Operations}: There are some simple math calculations happening, showing numbers being added or subtracted. This might be part of analyzing data or something similar.
  \item \textbf{Connection Details}: There are lines of numbers that represent sequence and acknowledgment numbers related to data being sent over a connection, as well as window size and length of data.
  \item \textbf{Frame Information}:
  \begin{itemize}[leftmargin=1em]
    \item A "Frame" refers to a packet of data intercepted during network communication, detailing its size (3517 bytes), where it was captured (on a network interface called eth0), and the source and destination of the packet.
    \item It includes the source and destination MAC addresses (unique identifiers for network devices) and IP addresses (used to identify devices on the internet).
  \end{itemize}
  \item \textbf{Protocol Information}:
  \begin{itemize}[leftmargin=1em]
    \item Mentions the Transmission Control Protocol (TCP), used for sending data between devices.
    \item Shows the source and destination ports (specific communication channels).
    \item Notes a large amount of data (3451 bytes) in the captured packet.
  \end{itemize}
  \item \textbf{Profile Information}: Indicates that the current setup is using the default settings of the program.
  \item \textbf{Statistics}: 1997 packets were analyzed and all were displayed (100\%).
  \item \textbf{Protocol Breakdown}: Mentions how the packets are divided by protocol types.
\end{itemize}
Overall, this text provides a technical snapshot of network traffic data captured by Wireshark. \\
\midrule
\textbf{Image-based input with multi-modal LLM} \\
\midrule
The image shows a screen from a program called Wireshark, which is used to analyze network traffic. Here's a simpler breakdown of its parts:
\begin{itemize}[leftmargin=*]
  \item \textbf{Display Filter}: A box where you can type to see only specific data or packets you want to examine.
  \item \textbf{Packets Seen}: This area lists all the data packets that have been captured. Each line shows details about a packet, like where it came from and what type it is.
  \item \textbf{Protocol Split of the Packets}: This section breaks down the packets by protocol — different languages or rules used for data transfer — and shows how many packets use each one.
  \item \textbf{Raw Packet}: Shows the raw data of a packet in both hexadecimal and ASCII format, providing a detailed look at the contents.
\end{itemize}
Overall, Wireshark helps users monitor and analyze the data being sent over a network. \\
\bottomrule
\end{tabularx}
\label{tab:slide4}
\end{table}

\begin{table}[hbt!]
\centering
\caption{LLM-assisted Instruction output for Figure \ref{fig:six-slides}(e)}
\small
\begin{tabularx}{\columnwidth}{>{\raggedright\arraybackslash}X}
\toprule
\textbf{Text-based input with OCR} \\
\midrule
Sure! Here’s a simpler version of the text:

\textbf{Instructions}

\textbf{Step 8: Move the ransomware to the target computer}
\begin{itemize}[leftmargin=*]
  \item Use a command to copy the ransomware file from your computer to the target computer that's on the same network. You'll use the terminal on your Attacker virtual machine (VM).
  \item Type this command: \texttt{scp ./ransomware ubuntu@\textless targetip\textgreater:~}, where \texttt{\textless targetip\textgreater} is the address of the target computer.
  \item When it asks for the password, enter \texttt{password} — this is the one you got from the previous step.
\end{itemize}

\textbf{Example in the terminal:}

You would see something like this: \\
\texttt{ubuntu@ip-10-100-12-221:~/AttackerDir\$ scp ./ransomware ubuntu@10.100.12.245:~} \\
\texttt{ubuntu@10.100.12.245's password:}

Then it will show that the ransomware file was copied successfully.

\textbf{Note}: This is a technical instruction for transferring a file and should be used responsibly. \\
\midrule
\textbf{Image-based input with multi-modal LLM} \\
\midrule
The image provides instructions for transferring a file called "ransomware" from one computer (the attacker’s virtual machine) to another computer (the target). Here's a simpler breakdown:
\begin{itemize}[leftmargin=*]
  \item \textbf{Step 8}: This step is about moving the "ransomware" file to the target computer.
  \item \textbf{Copy Command}: Use a command in the terminal to send the "ransomware" file over the network. It looks something like: \texttt{scp ./ransomware username@target-ip:~}, where \texttt{username} is the account name on the target computer, and \texttt{target-ip} is the address of the target computer.
  \item \textbf{Password}: When prompted for a password, enter the one you got earlier in the process.
\end{itemize}

This explains how to copy a file to another computer safely using command-line instructions. \\
\bottomrule
\end{tabularx}
\label{tab:slide5}
\end{table}

\begin{table}[hbt!]
\centering
\caption{LLM-assisted Instruction output for Figure \ref{fig:six-slides}(f)}
\small
\begin{tabularx}{\columnwidth}{>{\raggedright\arraybackslash}X}
\toprule
\textbf{Text-based input with OCR} \\
\midrule
This text describes the structure of a lab. Here's a simpler breakdown:
\begin{itemize}[leftmargin=*]
  \item \textbf{Module 1}: Learn how to capture Telnet passwords using a tool called Wireshark, which helps you see network data without actively interfering.
  \item \textbf{Module 2}: Learn how to capture SSH passwords, also using Wireshark.
  \item \textbf{Module 3}: Learn how to gather information actively using a tool called NMap, which scans networks to find details about devices and services.
\end{itemize}

This is part of a program at the University of Arizona. \\
\midrule
\textbf{Image-based input with multi-modal LLM} \\
\midrule
The image outlines a lab structure consisting of three parts:
\begin{itemize}[leftmargin=*]
  \item \textbf{Module 1}: Learn how to capture Telnet passwords using a tool called Wireshark, which records network data.
  \item \textbf{Module 2}: Learn how to capture SSH passwords, again using Wireshark.
  \item \textbf{Module 3}: Use a tool called NMap to actively gather information about network devices.
\end{itemize}

In simple terms, it's a workshop on how to collect passwords and information from networks using specific software tools. \\
\bottomrule
\end{tabularx}
\label{tab:slide6}
\end{table}

\end{document}